\documentstyle[prl,aps,preprint]{revtex}
\begin{document}
\newcommand{\lbf}[1]{\boldmath{\mbox{$#1$}}}
\draft
%\twocolumn[
%\hsize\textwidth\columnwidth\hsize\csname@twocolumnfalse\endcsname
\preprint{}
\title{
Joint effect of lattice interaction and potential fluctuation in colossal
magnetoresistive manganites
}
\author{R. Y. Gu} 
\address{
Texas Center for Superconductivity, University of Houston, Houston, TX 77204\\
and Ames Labortory, Iowa State University, Ames, IA 50011 
}
\date{\today}
\maketitle

\begin{abstract}
Taking into account both the Jahn-Teller lattice distortion and the on-site electronic potential fluctuations in the orbital-degenerated double-exchange model, in which both the core-spin and the lattice distortion are treated classically, we investigate theoretically the metal-insulator transition (MIT) in manganites by considering the electronic localization effect. An inverse matrix method is developed for calculation in which we use the inverse of the transfer matrix to obtain the localization length.
We find that within reasonable range of parameters, both the lattice effect and the potential fluctuation are responsible to the occurrence of the MIT. The role of the orbital configuration is also discussed.
\end{abstract}
%]

\newpage
\narrowtext

\section{introduction}
Colossal magnetoresistance (CMR), an effect discovered in mixed-valence manganites
R$_{1-x}$A$_x$MnO$_3$ (where R and A are rare and alkaline-earth ions, respectively) 
have attracted much attention \cite{kusters,jin,asamitsu,coey,varma}. In the hole doping range $0.17<x<0.4$, such an effect occurs in the vicinity of the
metal-insulator transition (MIT) temperature $T_{MI}$ at which the resistivity of 
the system varies drastically. It is generally accepted that the anomalous transport 
phenomena in these manganese systems are closely related to their magnetic properties,
in particular the paramagnetic (PM) to ferromagnetic (FM) phase transition upon
cooling, as in most experimental measurements, $T_{MI}$ and the magnetic Curie 
temperature $T_c$ are very close to each other.

The double exchange (DE) mechanism \cite{zener} is believed to play a major role in the
magnetic transition, in which the carriers are ferromagnetically coupled to the Mn
core spin due to a strong Hund's coupling, resulting in the hopping
amplitude depending on the relative Mn spin orientation. 
On the other hand, the mechanism responsible for the MIT has not yet been very clear \cite{coey2}. 
 It was found that in single orbital calculations the random hopping from the DE 
mechanism alone is not enough to drive the MIT \cite{sheng,li,calderon}.  
Many further theoretical efforts have been made to understand 
the MIT.  By taking into account 
the potential fluctuations experienced by carriers
due to the random distribution of $R^{3+}$ and $A^{2+}$, 
Sheng {\it et al.} \cite{sheng} proposed
that the electrons are localized above $T_c$ due to strong on-site and spin disorder, 
while at low temperature the alignment of the spins reduce the spin disorder and the 
electrons are
delocalized. In their calculation, however, the strength of the on-site disorder 
required for the localization,  $W\ge 12t$ (W is the range of the random on-site 
energy distribution, and $t$
is the one-orbital hopping amplitude), is apparently overestimated, as the reasonable magnitude of $t$ is roughly between 0.2 and 0.5 eV \cite{arima}, 
while the potential fluctuations would amount to
$1.7eV$ if unscreened, and be further reduced when screening is accounted \cite{coey}.
Millis {\it et al.} \cite{millis} used the
dynamical mean-field method to study the coupling of the carriers to local Jahn-Teller (JT)
distortions and to the Mn core spins, but an MIT was found only at half filling $(x=0)$. 
In recent years, there are also much interest on phase separation scenario 
\cite{yunoki,arovas,burgy},
in which the metallic and insulating regions coexist in the system and the transport is a percolation problem, near the percolation critical point, a small change of the fraction of the metallic regions can induce the MIT. But why this fraction change occurs with the variation of temperature still needs an explanation. Very recently, Verg\'{e}s {\it et al.} \cite{verges} proposed a lattice-spin mechanism to explain the MIT, by using Monto Carlo simulation in a $6^3$ lattice for a single-orbital DE model coupled with phonons, they found that at $x=0.08$, in a narrow region of the coupling parameter, an MIT coincidence with the FM to PM transition upon heating. 

Based on previous experimental and theoretical studies, it appears that the following
scattering mechanisms are important to the transport in manganites: 
(i) the strong DE interaction between carriers and the localized spin, which is usually
regarded as the basic mechanism in manganites, (ii) the 
potential fluctuations experienced by the electrons due to the
random $R^{3+}$ and $A^{2+}$ ion distribution, according to Ref.[\ref{coey}], this fluctuations
might be comparable to  the electron band-width, and (iii)
the JT distortion which was revealed in many experimental measurements.
Besides the basic DE interaction (i), in Refs.[\ref{sheng}] and [\ref{verges}], (ii) and (iii) were 
considered, respectively. Yet both theories have difficulties in their explanation 
of MIT (overestimated W in the former and not at the right doping in the latter).
Furthermore, in recent years it was also realized that 
in manganites the twofold $e_g$ orbital degeneracy and the unique 
Slater-Koster form of electronic hopping \cite{slater} are important to many aspects of the system \cite{brink}.
In many theoretical treatments, such as in Refs[\ref{sheng}] and [\ref{verges}], to simplify the calculation 
a single orbital model was adopted, as it was believed that the JT distortion would split the $e_g$ degeneracy and reduce the problem to an effective single orbital one.  
However, whether the energy splitting $\Delta$ is large enough to ignore the degeneracy is
questionable, in Ref.[\ref{verges}] the electron-lattice coupling responsible to the desired MIT is 
around $\lambda=1.45$, for such a strength of JT coupling, 
the average of the corresponding $\Delta$  is about 4(1-x)t, considerably smaller than the width of the 
energy band $12t$.  On the other hand, even in the $\Delta\rightarrow \infty$ limit where 
at each site there is only one orbital valid,
it is still different from the usual single orbital model due to the 
anistropy of the hopping, for example, in this large $\Delta$ limit if the  
lower-energy orbit is $d_{x^2-y^2}$, then along the z-direction there is no electronic hopping, 
very different from the isotropic hopping in the usual single orbital model. 

Because of the problems in previous MIT theories raised above, it is desirable to 
investigate the MIT of manganites by including  all the three scattering mechanisms 
as well as the orbital degeneracy. Up to now no such theoretical study has been made yet, and we will make 
such an effort in this work. Starting from an effective degenerated DE model with
electron-lattice interaction and potential fluctuation included, 
we investigate the MIT by considering the electronic localization effect.
An inverse matrix method suitable for the present specific Hamiltonian is developed for our calculation.  We find that while similar to the single orbital model,
in the degenerated orbital case the spin disorder alone is not enough
to localize the carriers either, the introduction of  the electron-lattice interaction and potential fluctuation, within reasonable regions of parameters,
can induce the MIT. We also discuss the role of orbital disorder in the MIT. It is shown that in the presence of strong electron-lattice interaction, electrons become more difficult to 
be localized in the orbital disordered state than in the orbital ordered state due to the split of orbital degeneracy and the anistropic electronic hopping in the latter.

\section{model and method of calculation}

We consider an effective DE Hamiltonian by taking into account the strong on-site 
Hund's coupling  between the itinant $e_g$ electrons and the Mn core 
spins, together with the coupling between electrons and lattice 
distortion,

\begin{equation}
H_e=-\sum_{ij}f_{ij}c^{\dag}_i\hat{t}_{ij}
c_j+\sum_i\epsilon_ic^{\dag}_ic_i
-g\sum_i Q_i c^{\dag}_i(\sin\phi_i\tau_x+\cos\phi_i
\tau_z)c_i+\frac{1}{2}\sum_{i}Q_i^2
\label{he}
\end{equation}
where $c_i^{\dag}=(c_{i+}^{\dag},c_{i-}^{\dag})$ with
$+$ and $-$ representing orbital states
$d_{3z^2-r^2}$ and $d_{x^2-y^2}$, respectively.
The first term of Eq. ($\ref{he}$) is the effective DE Hamiltonian in which
$f_{ij}=\cos(\theta_i/2)\cos(\theta_j/2)+\sin(\theta_i/2)
\sin(\theta_j/2)e^{-i(\varphi_i-\varphi_j)}$
with $(\theta_i,\varphi_i)$ being the polar angles characterizing the 
orientation of the localized spin on site $i$, 
$\hat{t}_{ij}=\hat{t}^{\alpha}$
are $2\times 2$ hopping matrix in orbital space, with
$\alpha(=x,y,z)$ being the directions of the bond (ij)
between neighboring Mn sites, and
$\hat{t}^{x,y}_{++}=t/4, \hat{t}^{x,y}_{+-}=\mp \sqrt{3}t/4$, 
$\hat{t}^{x,y}_{--}=3t/4$,
$\hat{t}^z_{++}=t, \hat{t}^z_{+-}=t^z_{--}=0$, and
$\hat{t}^{\alpha}_{-+}=\hat{t}^{\alpha}_{+-}$ \cite{yunoki}.
The second term represents the potential fluctuations experienced by $e_g$ electrons due to the
randomness of the $R^{3+}$ and $A^{2+}$ ion cores, for simplity we
assume that $\epsilon_i$ is uniformly distributed within the range [-W/2, W/2] \cite{sheng}.
The last two terms are related to the Jahn-Teller distortion,
in which $\tau_x, \tau_z$ are Pauli matrices and
$Q_i\sin\phi_i, Q_i\cos\phi_i$ are the two Jahn-Teller modes \cite{millis}.
Like in many other works, in Eq.(\ref{he}) we have used the classical limit for both the core spins and the Jahn-Teller distortions. In principle, the quantum lattice and spin fluctations could change some of the results.  In the case of large core spin $S=3/2$ and static lattice distortion, as is considered in this work, one expects that the quantum effect will not change the quanlitative conclusion of the classical approach.

In the following we will calculate the electronic localization length of Hamiltonian
(\ref{he}) in a long bar. Without losing generality we set the bar along 
the z direction, and the width of each slice of the bar is $M$. 
Usually the localization length $\lambda_M$ is obtained through the transfer matrix method \cite{mackinnon}, which is based on the recursion relationship of successive 
slices, $A_{n+1}=T_n^{-1}[(E-H_n)A_n-T_{n-1}A_{n-1}]$, where 
$A_n$ is the wave function amplitude in the $nth$ slice,
$E$ is the energy of electron, and $H_n$ is the $2M^2\times  2M^2$ 
matrix of the electronic Hamiltonian inside the slice, and 
$T_n$ is a $2M^2\times  2M^2$ diagonal matrix with elements being 
the electronic hopping amplitudes between neighboring sites 
in the $nth$ and $n+1th$ slices. The size of the corresponding transfer matrix 
is $4M^2\times 4M^2$. For the present specific Hamiltonian, 
this method can not be directly applied as  $T_n^{-1}$ does not exist 
due to the lack of hopping between neighboring  $d_{x^2-y^2}$ 
orbitals along the z direction.
The propagation of $d_{x^2-y^2}$ in the z direction is through its mixing 
with the $d_{3z^2-r^2}$ in the transverse directions. 
As a result, it is adequate to use the recursion equation of the 
latter to calculate the localization.  
By eliminating the amplitude of 
$d_{x^2-y^2}$ orbitals, the equation reduces to 
\begin{equation}
B_{n+1}=T_n^{\prime -1}[(E-H_n^{++}-U_n)B_n-T'_{n-1}B_{n-1}]
\label{recur2}
\end{equation}
where $B_n$ is the amplitude of $d_{3z^2-r^2}$ orbital,
$T'_n$ is the reduced $M^2\times M^2$ diagonal matrix whose  
$\nu$th diagonal elements is the effective 
hopping $f_{ij}$ between site $\nu$ ($\nu=1,2...,M^2$ denote the sites within the slice)
in the $nth$ slice and its neighboring in the $n+1th$ slice.  
$H_n^{\gamma\gamma'}$ $(\gamma, \gamma'=+, -)$ is the $M^2\times M^2$ submatrix of 
$H_n$ in subspace spanned by basis $\{c^{\dag}_{\nu\gamma}\}$ and $\{c_{\nu\gamma'}\}$,
$U_n=H_n^{+-}(E-H_n^{--})^{-1}H_n^{-+}$ 
is due to the mixing of the $d_{3z^2-r^2}$ and $d_{x^2-y^2}$ orbitals
inside the slice. 

The $2M^2\times 2M^2$ transfer matrix of each slice can be obtained 
from Eq.(\ref{recur2}). The localization length corresponds to the smallest 
positive Lyapunov exponent of the product of the transfer matrices, which comes
from, if we treat $B_n$ as a matrix, the behavior of its smallest eigenvalue in the limit of $n\rightarrow\infty$. 
In this method, since in a direct computation of the product matrix
the information associated with the smallest positive
Lyapunov exponent will be lost when the ratio of the contribution
from the smallest positive Lyapunov exponent to that of the
largest Lyapunov exponent becomes comparable with the
machine accuracy, one has to orthonormalize the product matrix regularly
\cite{mackinnon}.
Instead of this method, here we develop an inverse matrix
method in which there is no need to do the orthonormalization. Since the
smallest eigenvalue of $B_n$ corresponds to the largest eigenvalue of 
$B_n^{-1}$, we seek the recursion formula  of $B_n^{-1}$.
From Eq.(\ref{recur2}), it is given by
%\begin{mathletters}
%\begin{eqnarray}
%B_{n+1}^{-1}&=&B_n^{-1}g_{n+1} \label{bn} \\
%g_{n+1}&=&[E-H_n^{++}-U_n-T'_{n-1}g_n]^{-1}T'_n \label{gn}
%\end{eqnarray} 
%\end{mathletters}
\begin{equation}
B_{n+1}^{-1}=B_n^{-1}p_{n+1} \label{bn} 
\end{equation}
where the matrix $p_{n+1}$ can be directly obtained from 
recursion $p_{n+1}=[E-H_n^{++}-U_n-T'_{n-1}p_n]^{-1}T'_n$.
The largest eigenvalue of $B_{n}^{-1}$ is related to the
localization length and it increases exponently with $n$, so $B_{n}^{-1}$
needs to be regularly (but not necessarily at each $n$) renormalized by divided 
by $b_n=\sqrt{\sum_{ij}|(B_n^{-1})_{ij}|^2}$. 
%(($B_n^{-1})_{ij}$ are elements of $B_n^{-1}$).
For a bar of length $L$, the localization length is given
by $\lambda=L/|\sum'_n \log(b_n)|$, in which the sum is over all n where $B_n^{-1}$ are
renormalized. The result of $\lambda$ is not sensitive to the initial conditions
so that $p_1$ and $B_1$ can be simply set as the unit matrix.

The present inverse matrix method has been compared with the transfer matrix method. 
For the usual localization problems such as those in Ref.[\ref{mackinnon}], both methods 
give the same results within accuracy. For the present Hamiltonian Eq.(\ref{he}), 
we find that the former is more efficient due to the smaller matrix size and no need to
orthonormalize in the calculation (the disadvantage of having to inverse the 
matrix becomes less serious since in the present problem the latter 
also needs to perform matrix inversion in $U_n$). Moreover, it is found that the result of the latter lacks stability, for example, the results can be totally different by using single or double precision calculations, this deficiency may come from $T_{n}^{\prime -1}$ for very small effective hopping. On the other hand, there is no such problem in the former method. The difference in stability between these two methods is the main reason that
we choose the inverse matrix method in our calculation.

\section{results and discussion}
\subsection{Pure DE effect}
Now let us first see the case of a pure DE model ($g=W=0$) 
with orbital degeneracy. In this case, for a 
perfect FM phase all electronic states are extended, with
the energy band from -3$t$ to 3$t$. 
For hole-doped manganites R$_{1-x}$A$_x$MnO$_3$ ($0\le x\le 1$), the Fermi energy
$\varepsilon_F$ is at the bottom of the band near $x=1$, increases 
with the decreasing of $x$, and reaches 
the center of the band at $x=0$. With the increase of the randomness 
of spin configuration, 
the effective electronic hopping and energy band width decreases. 
At the PM phase, if we treat $\cos\theta_i$ and $\varphi_i$ as independent variables that are uniformly distributed in regions $[-1,1]$ and $[-\pi,\pi]$, then the band bottom is at around $-2.5t$. 
Under this treatment, Fig. 1 shows the calculated rescaled localization length as a function of energy E with different width M of the bar. 
The length of the bar is taken to be $L=10^5$, which is enough for our calculation. All curves are crossed at a fixed point $E_c\approx -2.0t$, which corresponds to the Fermi energy at doping $x_c\approx 0.85$.  The localization length in the thermodynamical 
limit $M\rightarrow\infty$ can
be obtained through the one-parameter scaling theory, in which the 
rescaled localization length $\lambda_M/M$
can be fitted in terms of a universal function 
$\lambda_M(E)/M=f(M/\xi(E))$ \cite{mackinnon}, and the scaling parameter $\xi$ has two branches
(inset of Fig. 1). For $E<E_c$, $\lambda_M(E)/M$ decreases toward zero and 
$\xi$ is the localization length $\lambda_{\infty}(E)$ for an
infinite system,  so electrons are localized and have no contribution
to the conductivity.  The electron state becomes extended when $E>E_c$, 
where $\lambda_M/M$ increases with $M$ and $\lambda_\infty$ is infinite.
The present calculation indicates that in
a pure DE model with orbital degeneracy, at doping $x<0.85$ in the PM state the DE spin 
disorder is not enough to localize the carriers on the Fermi surface and the
system is metallic. This conclusion is not consistent with the experimental 
results, where the MIT usually occurs in the range $0.17<x<0.4$.  

\subsection{Joint effect}
Next we take the other two scattering mechanisms into account. 
At each site the JT distortion is described by two quantities $Q_i$ and $\phi_i$.
For the amplitudes of the distortion $Q_i$, 
by minimizing the energy their average can be obtained as
$\langle Q_i\rangle=(1-x)g$. 
To simplify our study in the following calculations we 
will replace all $Q_i$ with $\langle Q_i\rangle$.
For the angle $\phi_i$, it is closely related to the orbital polarization
on site $i$, and since the configuration of orbitals 
at different sites are disordered at hole concentration  $0.17<x<0.4$ where 
the FM metallic to PM insulator transition occurs,
we put $\phi_i$ as random variables with uniform
distribution between $-\pi$ and $\pi$. Fig. 2 shows the calculated phase diagram for
the PM and perfect FM
magnetic background at $x=0.2$  in the $g-W$ plane, the energy is taken to be the 
Fermi energy at each $g$ and $W$.
Here we are interested in the region between the two MIT lines $L_{\mbox{\small PM}}$ and $L_{\mbox{\small FM}}$, which is an insulator in the PM and
a metal in the FM case, respectively, so that an MIT will occur in this region
when the magnetic configuration is changed from FM to PM. From Fig. 2, at $g=1.6$, 
to obtain the MIT, the required strength of the on-site disorder is $W\approx 3.5t$,
much smaller than that in Ref.[\ref{sheng}] (where $W\ge 12t$), and is
a reasonable magnitude according to the estimation in Ref.[\ref{coey}]. 
Fig. 2 is the main result of this paper, based on it we propose that the FM metallic
to PM insulator transition is due to the joint effect of the DE disorder, the JT distortion and the on-site potential fluctuation.
It is worth mentioning that in the present calculation  we have neglected the fluctuation of the lattice distortion amplitude $Q_i$, 
which in principle  depends on the electron occupation state on each site and has a thermal fluctuation 
around $\langle Q_i\rangle$. Since the on-site energy due to 
the lattice distortion at each site is proportional to $Q_i$, 
this fluctuation will introduce a potential randomness
in addition to that from $W$, so if considered it may further 
reduce the magnitude of $W$ required for localization.

\subsection{Effect of orbital configuration}
Since it appears that the 
on-site potential of orbitals $d_{3z^2-r^2}$ and 
$d_{x^2-y^2}$ (being $-gQ_i\cos\phi_i$ and $-gQ_i\sin\phi_i$, respectively) 
is more random in the orbital disordered case than in the ordered case and thus the diagonal disorder is stronger in the former, one might expect that  electrons are easier to be localized in the orbital disordered state  than in the orbital ordered state. To check this point we also perform calculations in the orbital ordered configurations and compare them with 
that in the orbital disordered case.
In Fig. 3  the rescaled localization length in both the orbital
disordered and C-type antiferromagnetic (AF) orbital ordered cases 
are shown, in the latter case
$\phi_i$ is taken to be $\pm 2/3\pi$ in the two sublattices,  
corresponding to the $d_{3x^2-r^2}/d_{3y^2-r^2}$ type order.
It is found that contrary to this expection,
electrons in the orbital ordered case electrons are actually easier to be localized. Similar results are also obtained for the A-type AF, G-type AF and FM orbital ordered configurations. This finding is consistent with the 
occurence of the orbital ordered insulator and orbital disordered metal 
observed in experiments\cite{endoh}.
In fact, for each
$\phi_i$, one can perform a local unitary transformation $U_i$ to transform the orbital state basis from $|+\rangle$, $|-\rangle$ to  
$|+'\rangle_i=\cos(\phi_i/2)|+\rangle+\sin(\phi_i/2)|-\rangle$ and 
$|-'\rangle_i=-\sin(\phi_i/2)|+\rangle+\cos(\phi_i/2)|-\rangle$. 
The electron-lattice interaction becomes diagonal under this transformation,
with the on-site energy being  $-gQ_i$ for
state $|+'\rangle_i$ and $gQ_i$ for $|-'\rangle_i$, respectively, which are 
independent of the orbital polar angle $\phi_i$. At the same time 
the hopping term in Eq.(\ref{he}) is transformed to $f_{ij}U_i^{\dag}\hat{t}_{ij}U_j$, so orbital disorder actually does not increase the randomness of the on-site potential  fluctuation, it only affects  that of the electronic hopping, which is already random in the PM phase due to the spin disorder. To our understanding, the localization effect of the
electron-lattice interaction comes from its
splitting the energy degeneracy. In the case of energy degeneracy, both 
orbital states $|+'\rangle_i$ and $|-'\rangle_i$ are important to
electronic transport and electron at a site can hop to either
the $|+'\rangle$ or the $|-'\rangle$ orbital state on its neighboring site. 
The electron-lattice interaction introduces an energy gap $\Delta$ between these two states.  With the increase of $\Delta$, it becomes more difficult for an electron at $|+'\rangle$ state to move to the neighboring 
$|-'\rangle$ state due to their energy difference, so that
electrons are easier to be localized in this case. 
To understand that the localization effect is even stronger in the
orbital ordered case, one may note that
at large $\Delta$, electron hopping occurs mainly
between neighboring $|+'\rangle$ states, whose hopping amplitude is
proportional to $\cos[(\phi_i-\phi_\alpha)/2]\cos[(\phi_j-\phi_\alpha)/2]$ 
($\alpha=x,y,z$ labels the (ij) bond direction, $\phi_{x,y}=\mp 2\pi/3$ and
$\phi_z=0$). In the orbital disordered case this 
amplitude fluctuated from bond to bond and the macroscopic transport
are isotropic. When orbitals are ordered, however, since
orbital ordered state is always accompanied by a global anistropy \cite{gu},
there is one direction along which the hopping amplitude is much smaller than that along the other direction(s), 
for example, in the FM orbital ordered case $\phi_i\equiv\phi$,
one can show that whatever the magnitude of $\phi$ is, there is always one direction along which the hopping integral is smaller than one quarter of that along another direction. In the special case of $\phi=\pi$ (corresponding to orbital $d_{x^2-y^2}$), along the z direction the hopping even vanishes. It is the weaker propagation along this
direction that enhances the localization effect of the whole system.  
 
\subsection{Temperature variation}
Finally let us briefly see the temperature induced MIT. 
In a given system the electron-lattice coupling $g$ and potential 
fluctuation $W$ are fixed while the spin disorder is temperature 
dependent. For our purpose the 
temperature dependence of the localized spin orientations is needed. 
Here we use the mean-field distribution of the orientations,  which
satisfies the self-consistent equation
$f(\theta_i,T)=F\exp(3\langle\cos\theta_i
\rangle\cos\theta_iT_c/T)$.  In Fig.4 we show the localization length as a 
function of the temperature.
For fixed parameters $g=1.7$ and $W=4t$, it is found that the MIT temperature 
$T_{MI}\approx 0.9T_c$ is close to, but not exactly at the 
Curie temperature $T_c$. Note that here 
the judgement of M or I phase is whether electrons are localized,
different from that in some other references where it is based on
the sign of $d\rho/dT$ so that the MIT temperature is always at the 
resistivity peak. Above $T_{MI}$ and below $T_c$, the localization length
decreases rapidly with the increase of the temperature due to the
sharp drop of the magnetization, as seen from Fig. 3, 
so that the resistivity can still increase
in this temperature region where transport is dominated by variable-range hopping 
of electrons \cite{varma}.
 
\section{Summary}
In summary, by using numerical scaling calculations for the localization effect in the orbital degenerated manganese system, we have shown that within reasonable range of parameters the MIT in manganites can be driven by the joint effect of the DE spin disorder, the on-site electronic potential fluctuation, and the JT distortion. 
We argue that all these three interactions are relevant to the FM metallic to PM insulate
transition in colossal magnetoresistive manganites.

\acknowledgements{
This work was supported by the Texas Center for Superconductivity at the 
University of Houston and the Robert A. Welch Foundation,
and was also supported in part by the Ames Laboratory, which is operated by 
Iowa State University for the U.S. Department of Energy (DOE) under Contract No. W-7405-ENG-82. 
}

\begin{figure}
\caption{Rescaled localization length as a function of the energy $E$. The inset is the scaling parameter $\xi$ as a function of $E$. The length of the bar is taken to be $L=10^5$. 
}
\end{figure}

\begin{figure}
\caption{
Phase diagram in the PM and FM magnetic background at $x=0.2$. 
$L_{\mbox{\small PM}}$ and $L_{\mbox{\small FM}}$ are MIT lines in the PM and FM states, respectively. 
Regions below $L_{\mbox{PM}}$ and above $L_{\mbox{FM}}$ are 
metal and insulator in both the PM and FM cases, while the region in between is insulator in the PM state and metal in the FM state. 
}
\end{figure}

\begin{figure}
\caption{Comparison of the
localization lengths in the orbital disordered (open symbols) and C-type AF ordered (solid symbols) configurations in the PM state at $g=1.4$ and $W=4t$.
}
\end{figure}

\begin{figure}
\caption{Temperature dependence of the localization lengths at $g=1.7$ and $W=4t$.
}
\end{figure}

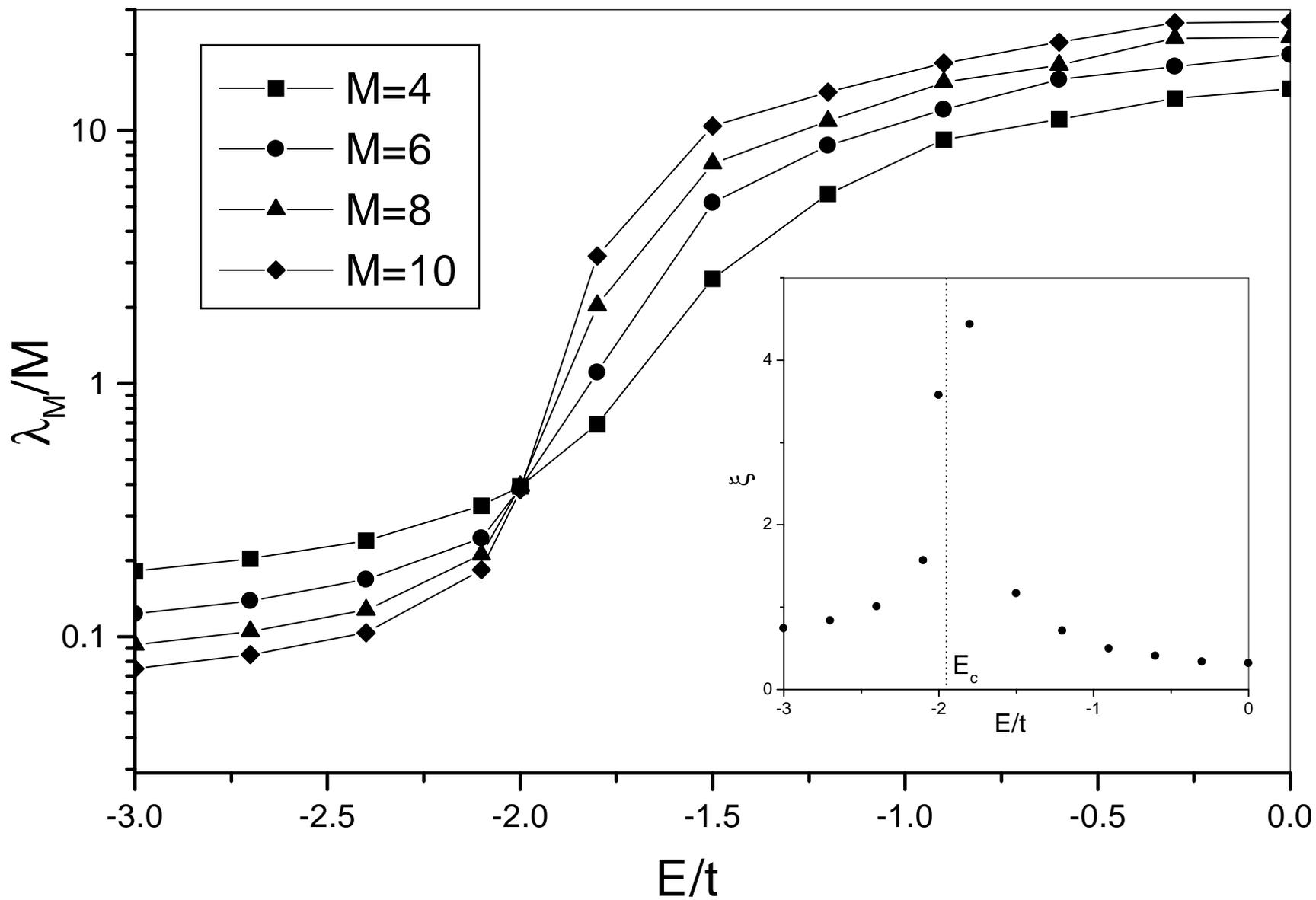
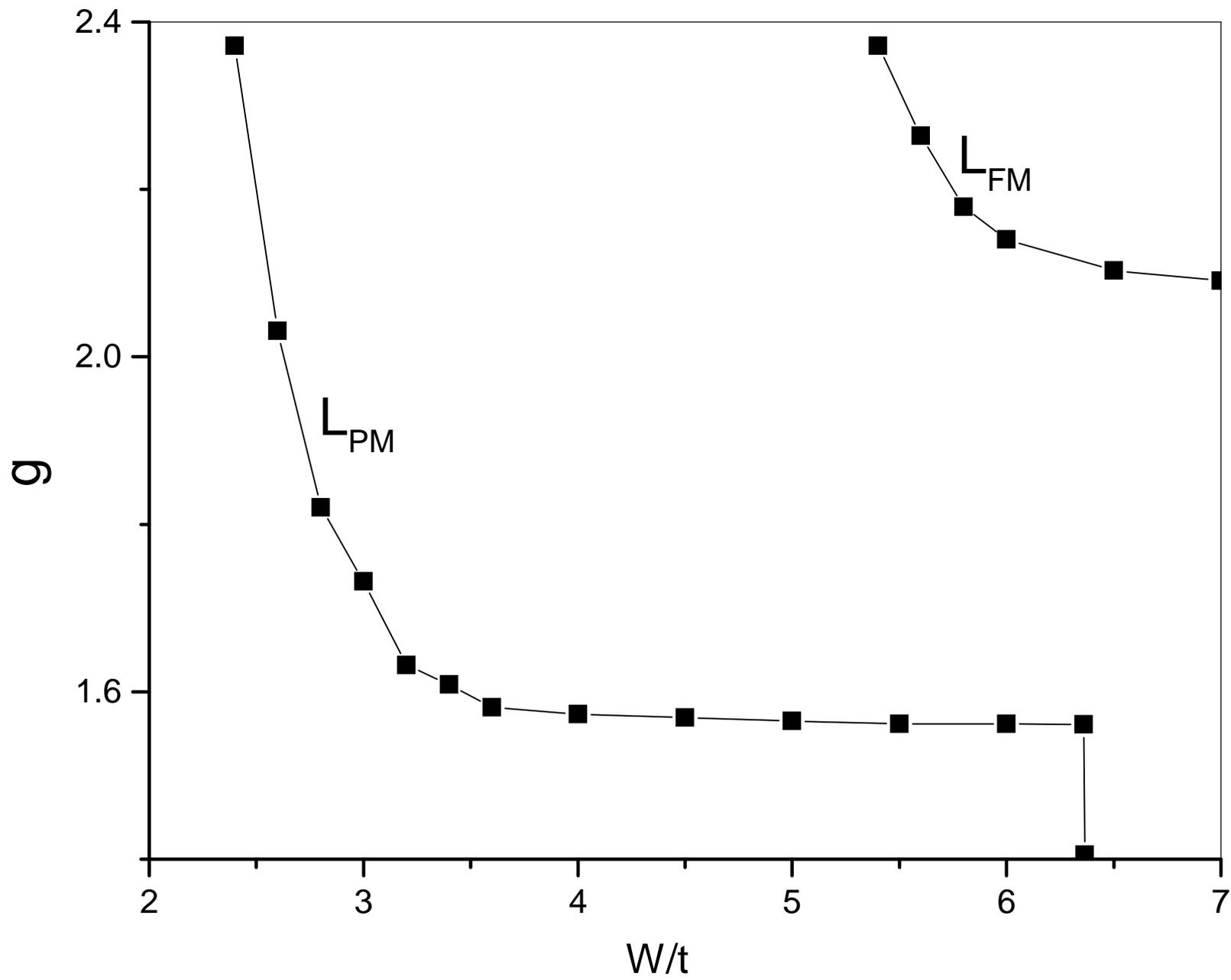
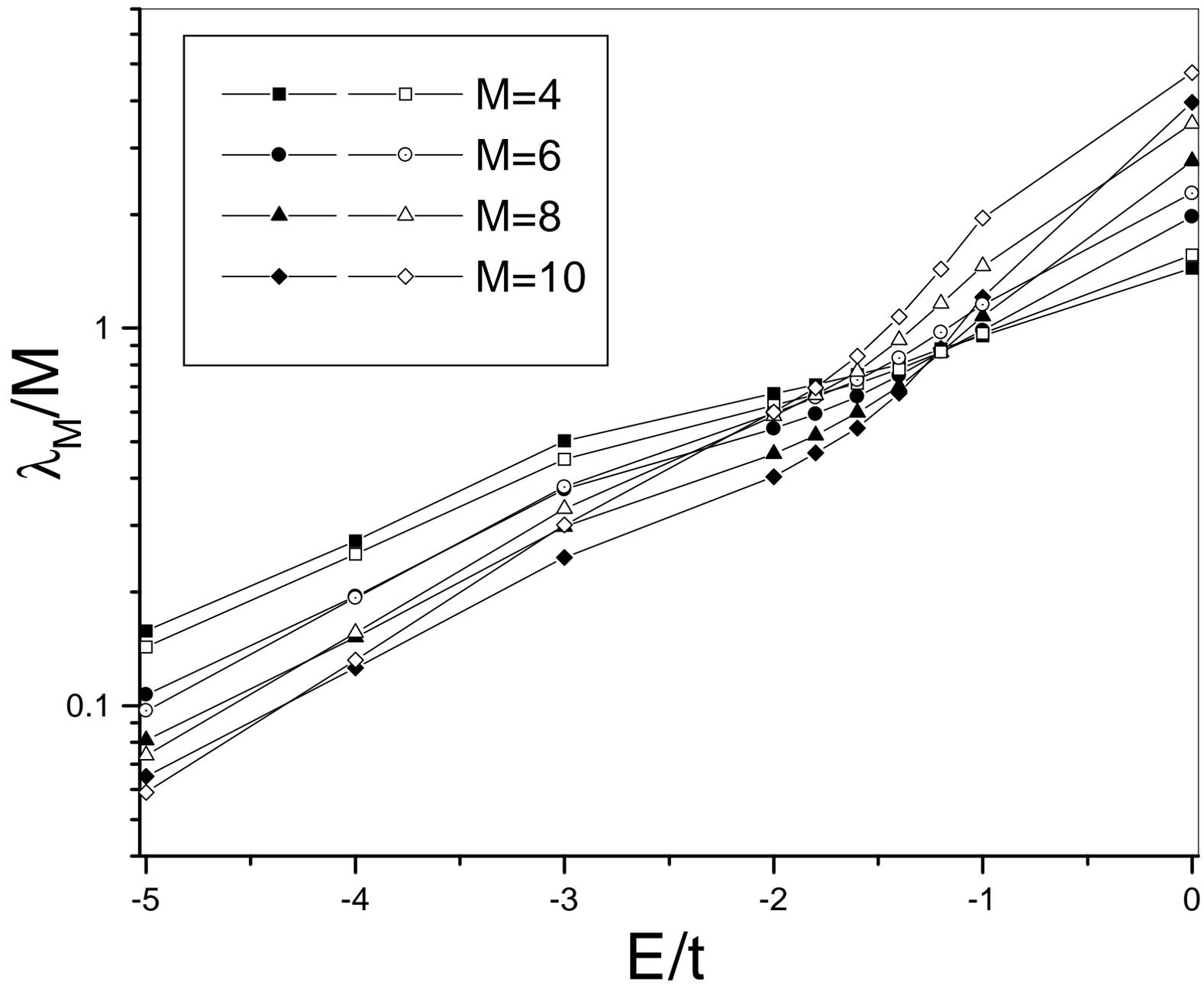
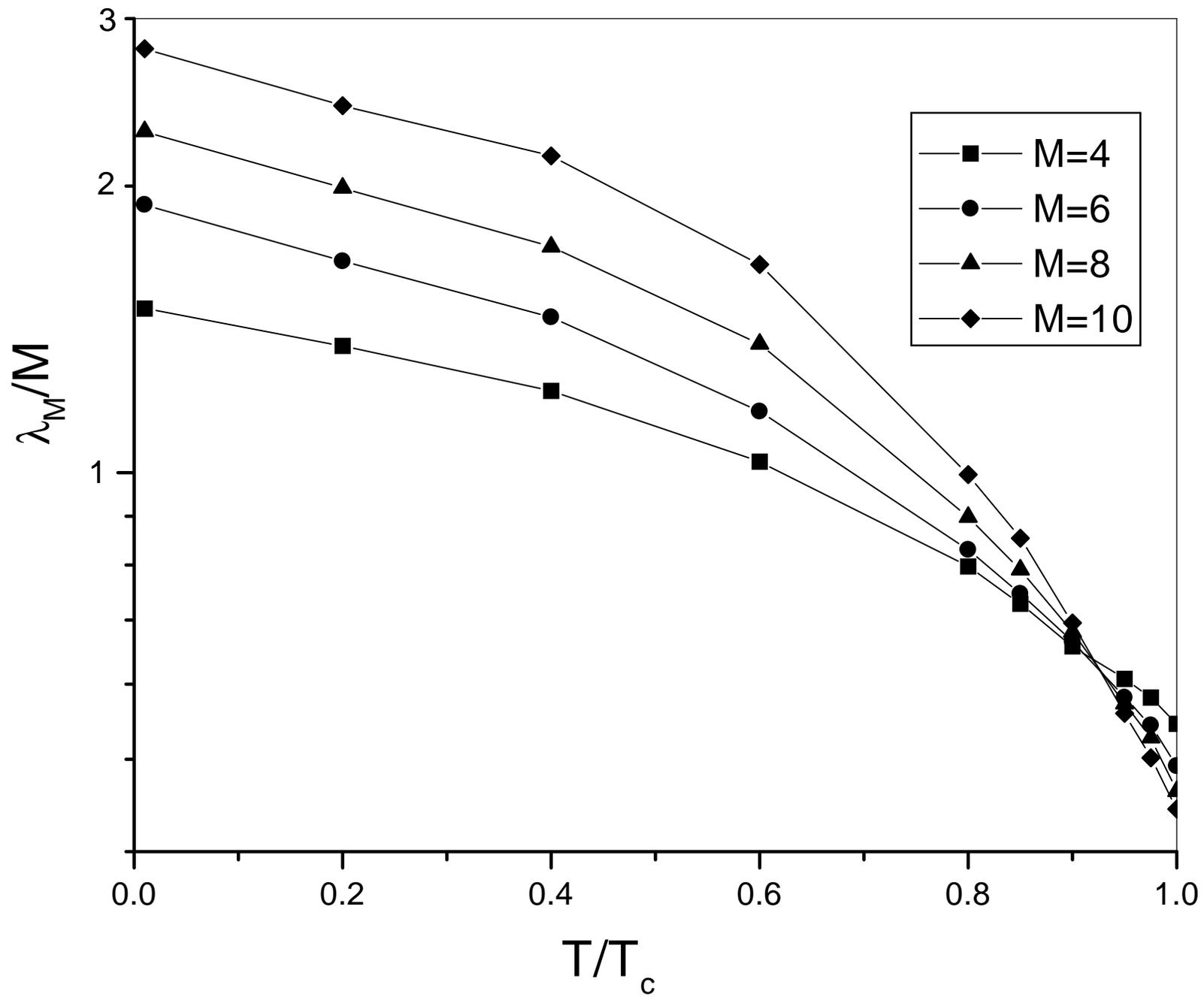
\end{document}